\documentclass[a4paper,11pt]{article}
\usepackage{pos}
\usepackage{float}  
\usepackage{subcaption} 
\usepackage[dvipsnames]{xcolor}

\usepackage{comment}
\usepackage{booktabs}
\usepackage{multirow}
\graphicspath{{./figures/}}

\title{HLbL contribution to the muon g-2 using twisted-mass fermions at the physical point }

\author*[a]{N.~Kalntis}
\author[b]{G.~Kanwar}
\author[c]{M.~Petschlies}
\author[a]{S.~Romiti}
\author[a]{U.~Wenger}

\affiliation[a]{Institute for Theoretical Physics, Albert Einstein Center for Fundamental Physics, University of Bern, CH-3012 Bern, Switzerland}
\affiliation[b]{Higgs Centre for Theoretical Physics, University of Edinburgh, Edinburgh EH9 3FD, UK}
\affiliation[c]{Helmholtz-Institut f{\"u}r Strahlen- und Kernphysik (Theorie), University of Bonn, 53115 Bonn, Germany}
\emailAdd{nikolaos.kalntis@unibe.ch}

\abstract{We present updated results for the hadronic light-by-light (HLbL) contribution to the muon anomalous magnetic moment. The calculations are based on ETMC's $N_f=2+1+1$ Wilson-clover twisted-mass ensembles at the physical point. We perform continuum extrapolations for the strange- and charm-quark connected contributions and report on our results for the light-quark connected and light-quark 2+2 disconnected contributions at one lattice spacing.
\begin{center}
\includegraphics[width=0.3\textwidth]{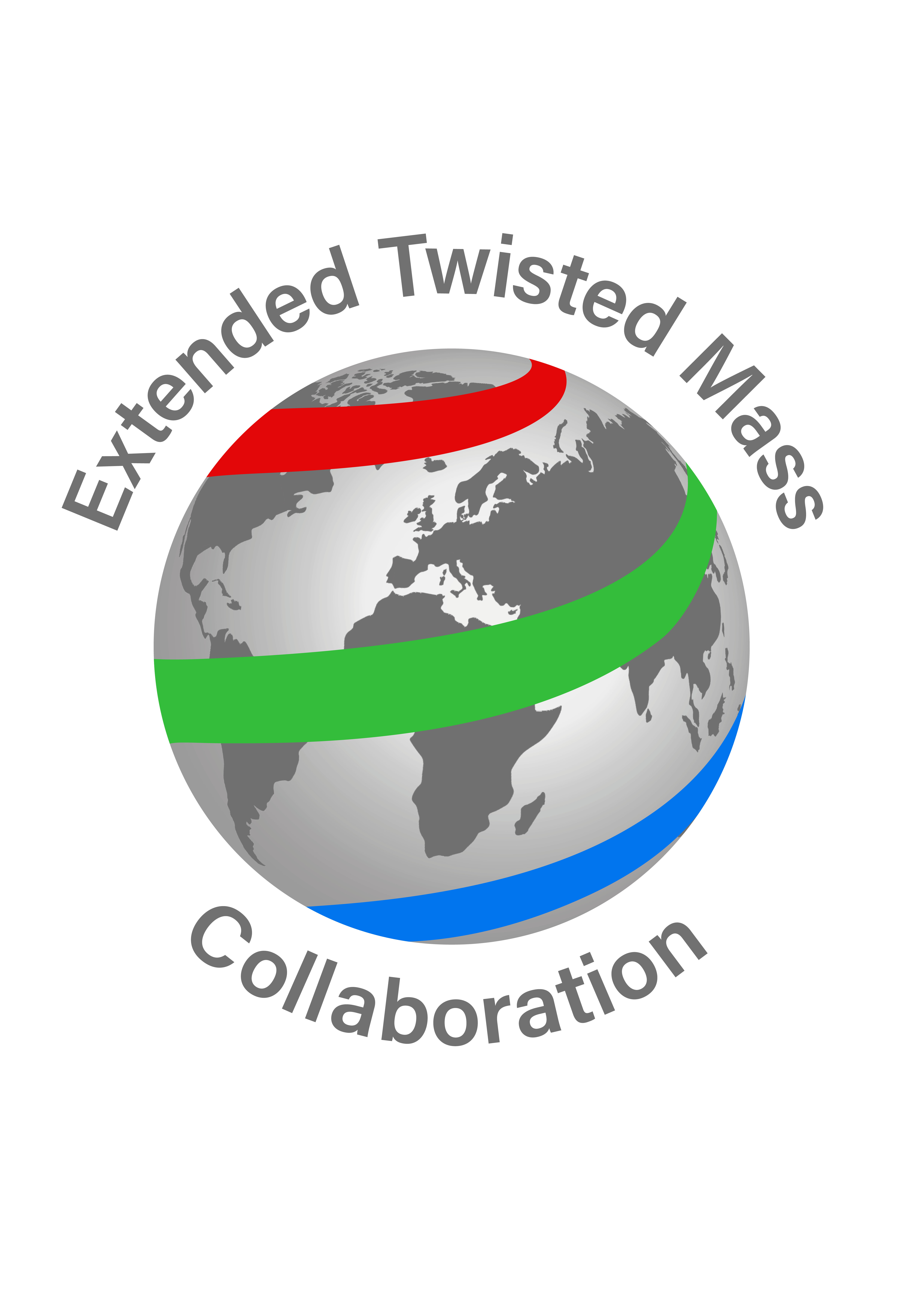}
\end{center}
}

\FullConference{The 42nd International Symposium on Lattice Field Theory (LATTICE2025)\\
2-8 November 2025\\
Tata Institute of Fundamental Research, Mumbai, India\\}

\begin{document}
\maketitle

\section{Theoretical background}

We report on our ongoing calculation of the hadronic light-by-light (HLbL) contribution to the anomalous magnetic moment of the muon $a^{\mathrm{HLbL}}_\mu$ from lattice QCD. For this computation we use $N_f=2+1+1$ twisted-mass clover-improved fermions at maximal twist, which leads to automatic $\mathcal{O}(\text{a})$ improvement on the calculated observables \cite{Frezzotti:2003ni, Frezzotti:2004wz}. The gauge ensembles are generated by the Extended Twisted Mass Collaboration (ETMC) \cite{ExtendedTwistedMass:2021gbo,ExtendedTwistedMass:2022jpw}.  The quark masses for these ensembles are tuned such that the charged-pion mass is fixed to its physical value and
the $s$- and $c$-quark masses are tuned to reproduce the physical $K$ and $D_s$ meson masses as in Ref.~\cite{ExtendedTwistedMass:2024nyi}. The details of these ensembles are summarized in Table \ref{tab:ensemble_details}.
\begin{table}[b]
\small
\centering
\[
\begin{array}{|c||c|c|c|c|c|c|c|}
\hline
\text{Ensemble} & L^3 \cdot T /a^4 & M_{\pi}~[\text{MeV}] & a~[\text{fm}] & L~ [\text{fm}] & M_\pi L & Z_V \\ \hline\hline
\text{cB64} & 64^3 \cdot 128 & 140 & 0.07951(4) & 5.09 & 3.6 & 0.706377(20)  
\\ \hline
\text{cC80} & 80^3 \cdot 160 & 137 & 0.06816(8)  & 5.45 & 3.8 & 0.725405(14) 
\\ \hline
\text{cD96} & 96^3 \cdot 192 & 141 & 0.05688(6) & 5.46 & 3.9 & 0.744110(7)   
\\ \hline
\text{cE112} & 112^3 \cdot 224 & 136 & 0.04891(6)  & 5.48 & 3.8 & 0.758231(5) \\ \hline
\end{array}
\]
\caption{Properties of the ensembles used for this work; further details are available in Refs.~\cite{ExtendedTwistedMass:2024nyi,ExtendedTwistedMass:2024myu}.}
\label{tab:ensemble_details}
\end{table}
Compared to our previous results in Ref.~\cite{Kalntis:2024dyd}, we have added a finer ensemble (cE112) for improving the continuum extrapolation of the heavy-quark contributions, while we compute the light-quark contributions with improved statistics compared to our previous calculation.

We follow the position-space approach put forward by the Mainz collaboration \cite{Green:2015mva, Asmussen:2016lse, Chao:2020kwq, Chao:2021tvp,Chao:2022xzg}, where the HLbL contribution to $a_\mu$ can be calculated as 
\begin{equation} 
    a_\mu^{\text{HLbL}} = \frac{me^6}{3} \int d^4 x \, d^4 y~\Bar{\mathcal{L}}^{(\Lambda)}_{[\rho,\sigma];\mu\nu\lambda}(x,y)~i \Hat{\Pi}_{\rho;\mu\nu\lambda\sigma}(x,y) \, . 
    \label{eq:master_formula}
\end{equation}
In Eq.~\eqref{eq:master_formula}, the modified QED kernel $\Bar{\mathcal{L}}^{(\Lambda)}$ is a known semi-analytic function associated with the perturbative QED part \cite{Asmussen:2022oql}, while $\Hat{\Pi}$ is the hadronic four-point function calculated on the lattice. The four-point function is defined by

%
\begin{equation}
i\Hat{\Pi}_{\rho;\mu\nu\lambda\sigma}(x,y) = - \int d^4z\, z_\rho 
\langle j_\mu(x) j_\nu(y) j_\sigma(z) j_\lambda(0) \rangle_{\text{QCD}}\,,
\label{eq:Pi_hat_combined}
\end{equation}
%
with $j_\mu(x) = \displaystyle\sum_{f = u, d, s, c} Q_f\, (\Bar{q}_f \gamma_\mu q_f)(x)$ the electromagnetic current and $Q_f$ the electromagnetic charges of the various quark flavors. 
One can use $O(4)$ rotational symmetry to exactly integrate out the angular components of $y$, rewriting Eq.~\eqref{eq:master_formula} as a function of $|y|$ only. Introducing an upper bound $|y|_{\mathrm{max}}$, we write
\begin{equation}
\begin{gathered}
a_\mu^{\mathrm{HLbL}} = \displaystyle\lim_{|y|_{\mathrm{max}} \to \infty} a_\mu^{\text{HLbL}}(|y|_{\text{max}}), \quad
a_\mu^{\mathrm{HLbL}}(|y|_{\mathrm{max}}) \equiv \int_0^{|y|_{\text{max}}} d|y|\, f(|y|), \\
f(|y|) \equiv \frac{me^6}{3}\, 2\pi^2 |y|^3 \int d^4x\, 
\Bar{\mathcal{L}}^{(\Lambda)}_{[\rho,\sigma];\mu\nu\lambda}(x,y)\, i\Hat{\Pi}_{\rho;\mu\nu\lambda\sigma}(x,y)\,.
\end{gathered}
\label{eq:amu-fy}
\end{equation}
An arbitrary fixed choice of direction $y = |y| \hat{y}$ is sufficient to evaluate $f(|y|)$ in Eq.~\eqref{eq:amu-fy}. On the lattice we adopt the local version of the electromagnetic current (which we renormalize with $Z_V$ as in Table~\ref{tab:ensemble_details}), and
all spacetime integrals become sums over the lattice points. In our calculation, we perform an explicit summation over the variables $x$ and $z$,
while we select specific points $y$ corresponding to a certain range of $|y|$ and integrate the one-dimensional integral in Eq.~\eqref{eq:amu-fy} by quadrature. The function $f(|y|)$ is referred to in the following as the \textit{integrand}.

The evaluation of the QCD four-point function contains five classes of Wick contractions~\cite{Chao:2020kwq}. In this work we focus on the two dominant ones: the fully-connected diagram (for the $\ell,s,c$ quarks) and the 2+2 disconnected diagram (for the $\ell$ quarks).  
For the calculation of the charm-quark contribution we use the two modified kernels $\overline{\mathcal{L}}^{(3)}$ and $\overline{\mathcal{L}}^{(\Lambda=0.4)}$ as defined in Refs.~\cite{Asmussen:2019act, Chao:2020kwq},
in order to check for discretization and finite-volume effects. 
To further check for discretization effects and resolve better the integrand, we use two different directions $\hat{y}$; the diagonal ``1111'' direction, defined by $\hat{y}/a \propto (1,1,1,1)$, 
and the off-diagonal ``0111'' direction, defined by $\hat{y}/a \propto (0,1,1,1)$.
For the strange-quark contribution we use both kernels $\overline{\mathcal{L}}^{(3)}$ and $\overline{\mathcal{L}}^{(\Lambda=0.4)}$, but only the direction 1111. For the light-quark contributions, we use only the kernel $\overline{\mathcal{L}}^{(\Lambda=0.4)}$ and direction 1111. 



\section{Heavy-quark connected contributions}
\label{subsec:heavy_quarks}

In this section we show the results for the fully-connected contributions coming from the strange and charm quarks based on the statistics as given in Table~\ref{tab:StrangeCharmPlots}.

\begin{figure}[pth!]
    \centering
    \begin{subfigure}[b]{\textwidth}
        \centering
        \includegraphics[width=0.9\textwidth]{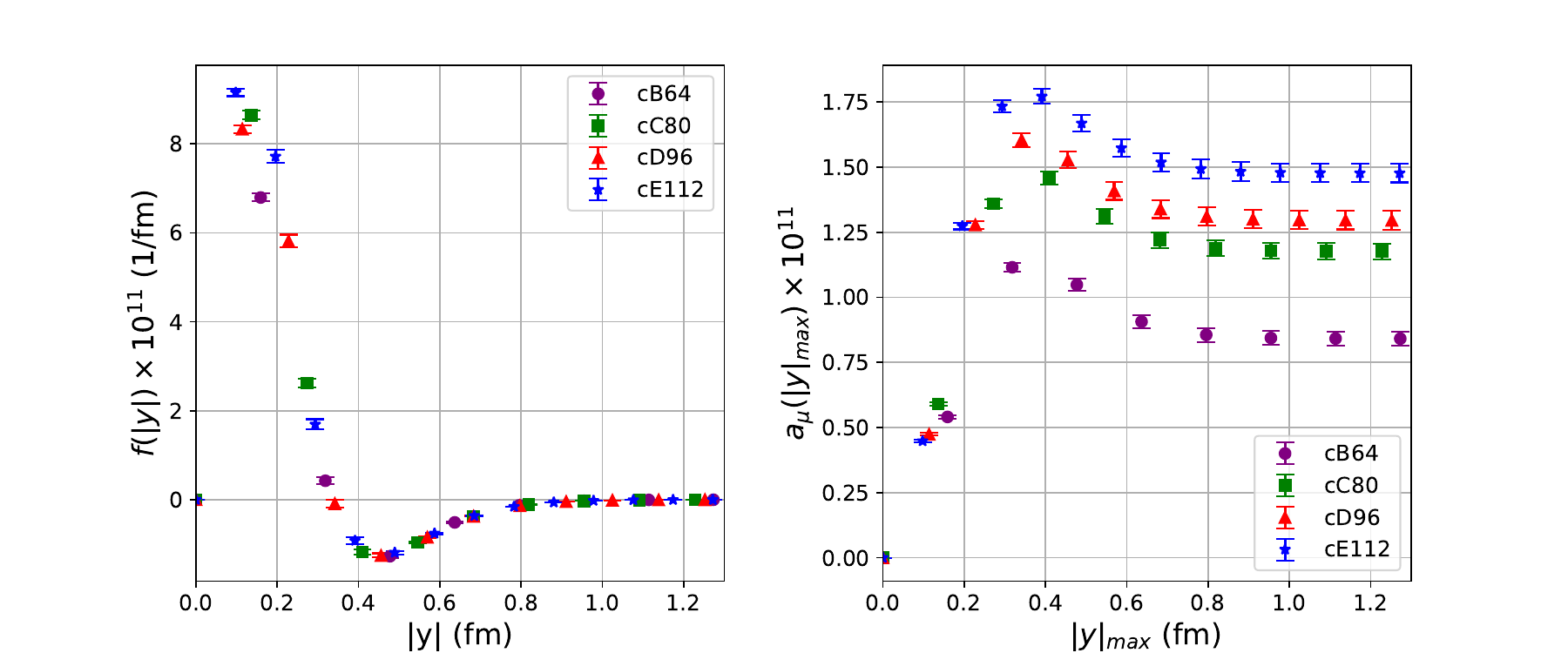}
        \label{fig:charm-k4-1111}
    \end{subfigure}
    
    \begin{subfigure}[b]{\textwidth}
        \centering
        \includegraphics[width=0.9\textwidth]{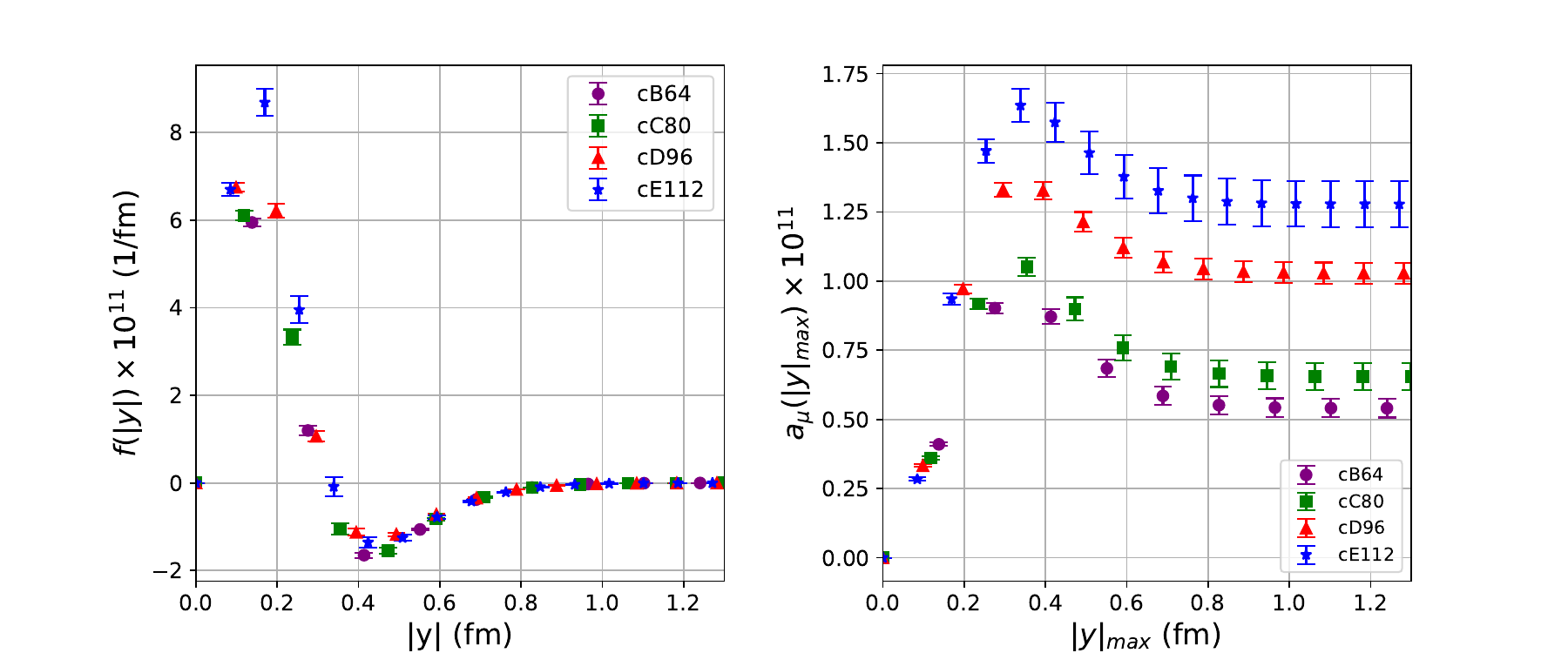}
        \label{fig:charm-k4-0111}
    \end{subfigure}
    
    \caption{Results for the charm-quark connected contribution evaluated on the ensembles cB64, cC80, cD96 and cE112 for kernel $\overline{\mathcal{L}}^{(\Lambda=0.4)}$ for the 1111 direction (upper two plots) and 0111 direction (lower two plots). The left plots depict the integrand $f(|y|)$ as a function of the distance $|y|$ and the right plots the partially integrated $a_\mu(|y|_\text{max})$ as a function of the cut-off distance $|y|_\text{max}$.}
    \label{fig:charm-kernels}
    \centering
    \vspace{0.25cm}
    \includegraphics[width=0.8\textwidth]{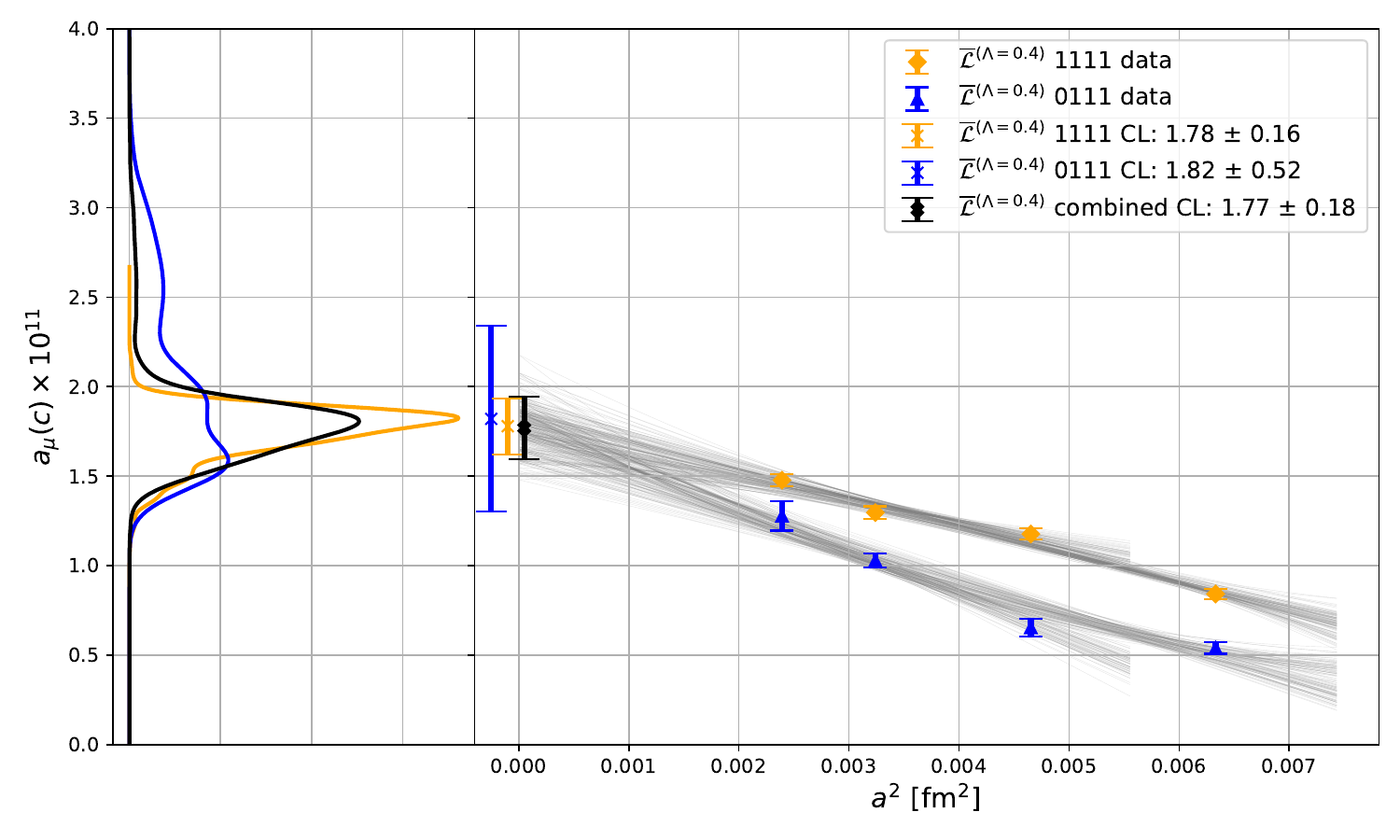}
    \caption{Continuum extrapolation (CL) of the charm-quark connected contribution for kernel $\overline{\mathcal{L}}^{(\Lambda=0.4)}$. 
    Gray curves show a subset of all fits considered, with opacity indicating their relative AIC weights. The left panel shows the AIC-weighted distribution of central values.}
    \label{fig:continuum-charm-connected}
\end{figure}

The results for the charm-quark connected contribution are shown in Figures \ref{fig:charm-kernels} and \ref{fig:continuum-charm-connected}. In order to determine the continuum-extrapolated value and account for systematic effects, we employ the Akaike Information Criterion (AIC), following Ref.~\cite{Borsanyi:2020mff}. In the AIC analysis, we include fits over all four 
or only the three finest lattice spacings, using linear and quadratic dependence in $a^2$, and we impose a common continuum limit for the 1111 and 0111 directions.

\begin{figure}[pt]
    \centering
    \begin{subfigure}[b]{\textwidth}
        \centering
        \includegraphics[width=\textwidth]{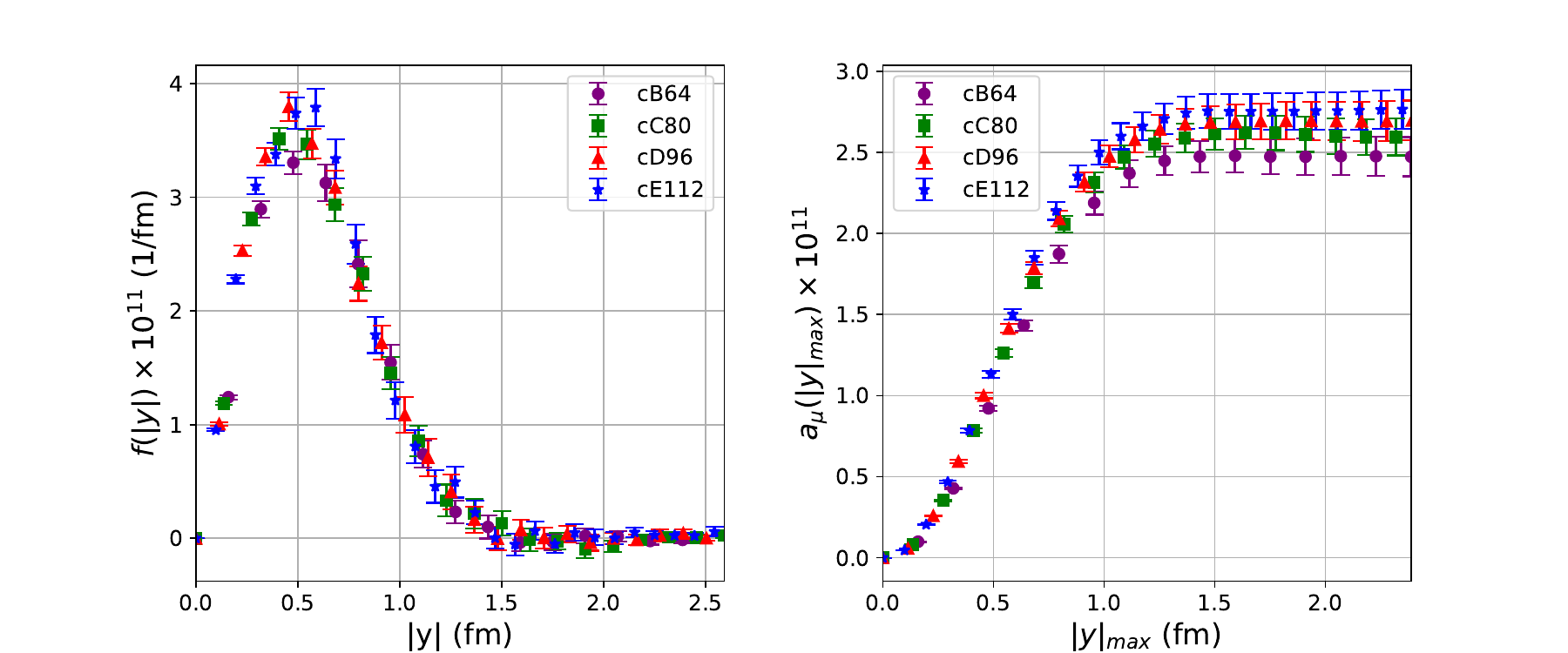}
        \label{fig:strange-k4}
    \end{subfigure}
    
    \caption{Results for the strange-quark connected contribution on the ensembles cB64, cC80, cD96 and cE112 for kernel $\overline{\mathcal{L}}^{(\Lambda=0.4)}$ and diagonal direction 1111. The left plot depicts the integrand $f(|y|)$ as a function of the distance $|y|$ and the right plot the partially integrated $a_\mu(|y|_\text{max})$ as a function of the cut-off distance $|y|_\text{max}$.}
    \label{fig:strange-kernels}
    \centering
    \vspace{0.5cm} 
    \includegraphics[width=0.8\textwidth]{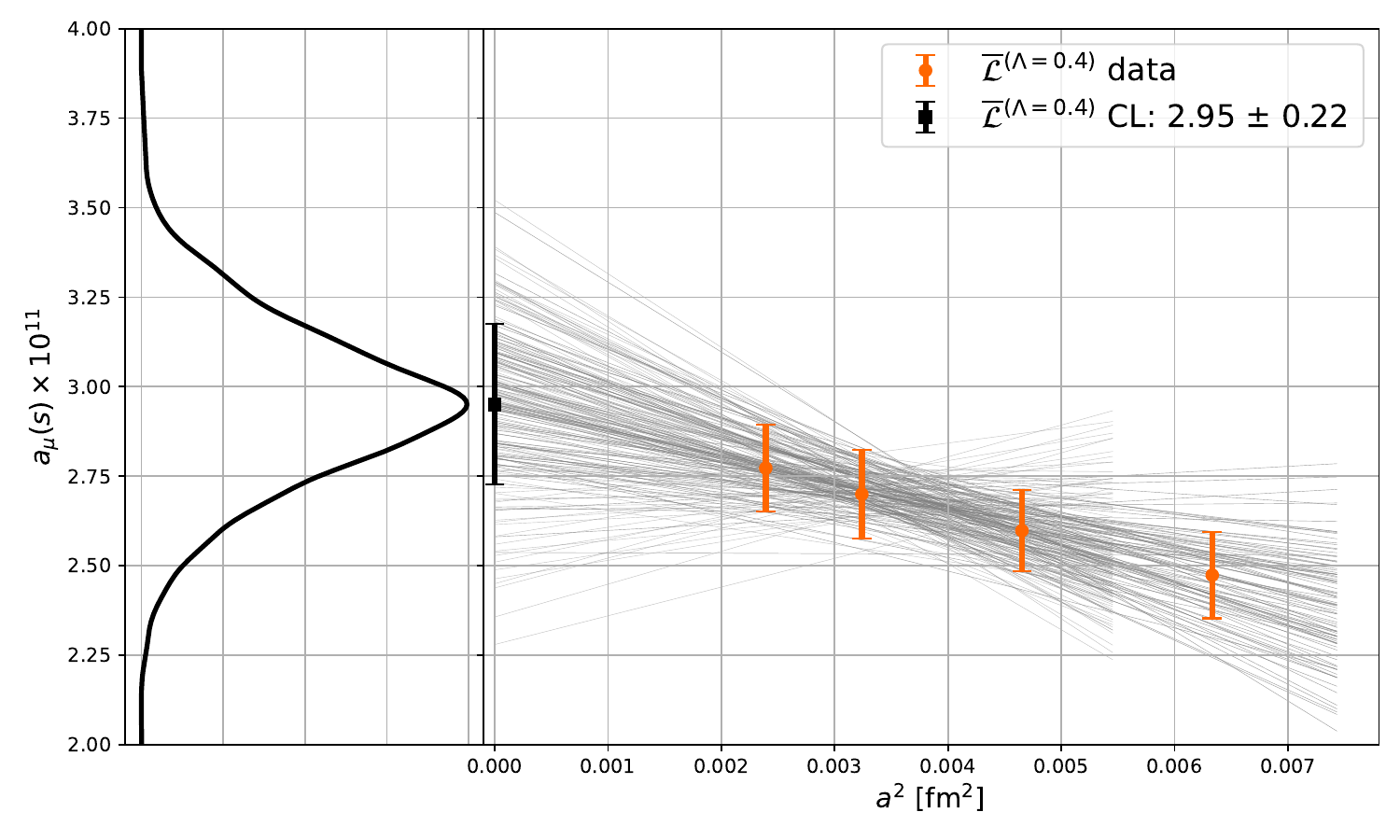}
    \caption{Continuum extrapolation (CL) of the strange-quark connected contribution for kernel $\overline{\mathcal{L}}^{(\Lambda=0.4)}$. Gray curves show a subset of all fits considered, with opacity indicating their relative AIC weights. The left panel shows the AIC distribution of central values.}
    \label{fig:continuum-strange-connected}
\end{figure}

The results for the strange-quark connected contribution, this time only for the diagonal direction 1111, are shown in Figs.~\ref{fig:strange-kernels} and \ref{fig:continuum-strange-connected}. For the continuum extrapolation we again follow the same AIC procedure as in the charm case, this time including however only linear fits in $a^2$ with either all four or only the three finest lattice spacings.  We do not include terms quadratic in $a^2$, since the corresponding coefficients are compatible with zero and have a too large uncertainty. 
The lattice artifacts for the strange-quark connected contribution are smaller compared to the charm-quark connected contrinution, as indicated by the flatter continuum extrapolations.

\begin{table}[b]
    \centering
    \begin{tabular}{ccccc}
        \toprule
         Flavor & Configs. & C.L.\ $\Bar{\mathcal{L}}^{(3)} \times 10^{11}$ & C.L.\ $\Bar{\mathcal{L}}^{(\Lambda=0.4)} \times 10^{11}$ & Plots \\
         \midrule
         Charm  & (1111): 260, (0111): 60 & 1.72(19) & 1.77(18) & Figures~\ref{fig:charm-kernels}, \ref{fig:continuum-charm-connected}
         \\
         Strange & (1111): 260  &  2.70(26) & 2.95(22) &  Figures~\ref{fig:strange-kernels}, \ref{fig:continuum-strange-connected}
        \\
         \bottomrule
         \end{tabular}
         \caption{Table of preliminary results for the heavy-flavor contributions from the fully-connected diagrams. We report the number of configurations (same for each ensemble) used to evaluate the contributions for each choice of direction (as indicated in parentheses). The continuum limit (C.L.) values are shown for both choices of QED kernel, with total uncertainties obtained from the AIC averaging described in the main text.}
    
    \label{tab:StrangeCharmPlots}
\end{table}

The preliminary continuum-extrapolated values for the charm- and strange-quark connected contributions for both kernels, including all statistical and systematical effects (via the AIC procedure), are summarized in Table~\ref{tab:StrangeCharmPlots}.
The extrapolated values for kernel $\overline{\mathcal{L}}^{(3)}$ are consistent with those for the kernel $\overline{\mathcal{L}}^{(\Lambda = 0.4)}$ for both the strange- and charm-quark connected contribution.
For the latter, further investigation is under way in order to suppress lattice artifacts at short distances.  In the current picture there is significant discrepancy between the results presented in this report, the results from BMW \cite{Fodor:2024jyn} and the results from Mainz \cite{Chao:2021tvp}.

\section{Light-quark connected and 2+2 disconnected contributions}
\label{subsec:light_connected}

\begin{figure}[pt]
    \centering
    \includegraphics[width=\textwidth]{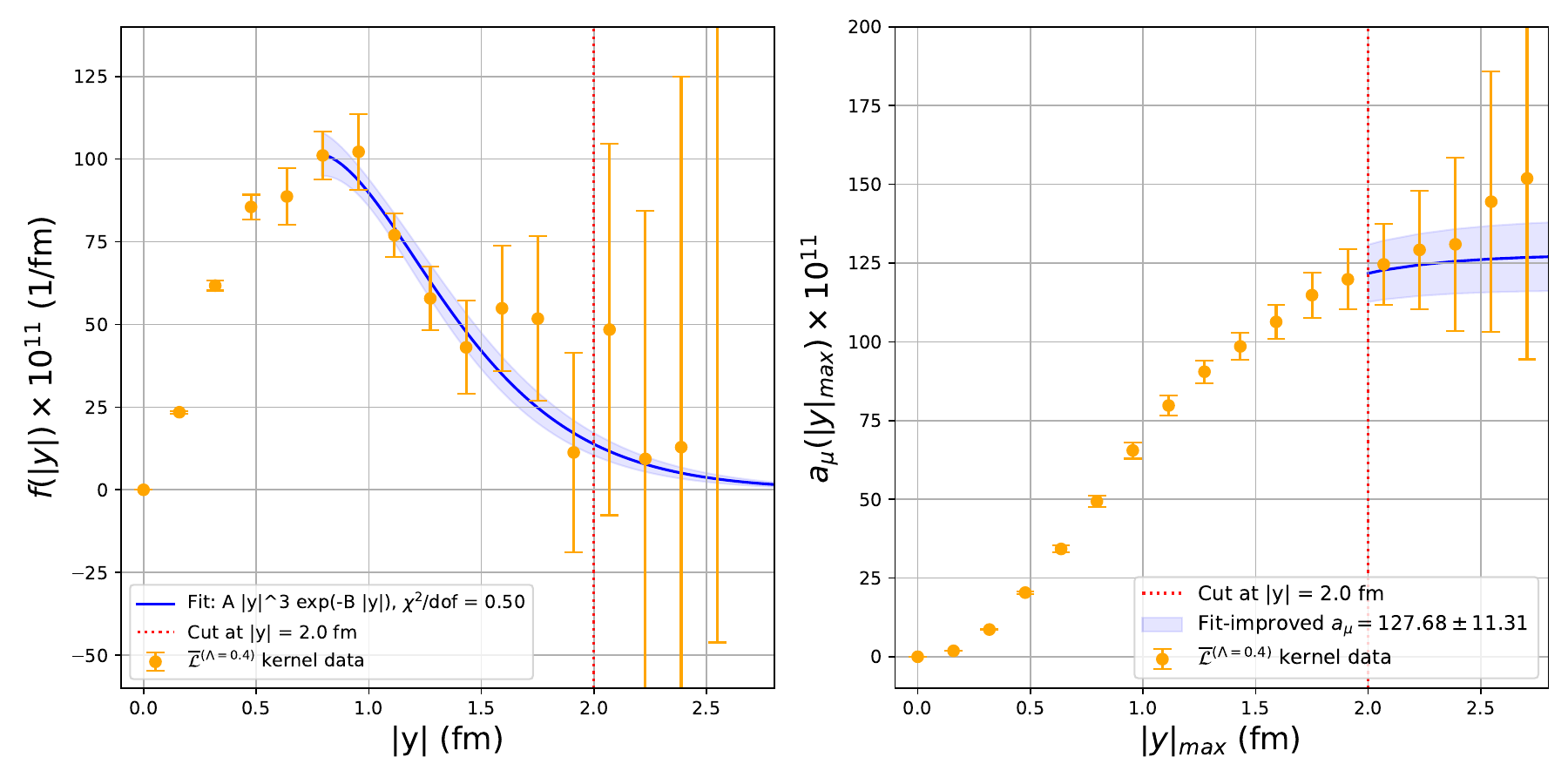}
    \caption{
    Preliminary results for the light-quark connected contribution on the ensemble cB64.
    The left plot depicts the integrand $f(|y|)$, with a fit of the form $A~|y|^3~e^{-B|y|}$ for $|y| \gtrsim 0.79$ fm shown as a blue band. The right plot displays the partially integrated $a_\mu(|y|_\text{max})$ with the effect of replacing the data by the fit result for $|y| \geq 2.0$ fm shown as a blue band.
    }
    \label{fig:light-connected-k4}
    \centering
    \vspace{0.5cm}
    \includegraphics[width=\textwidth]{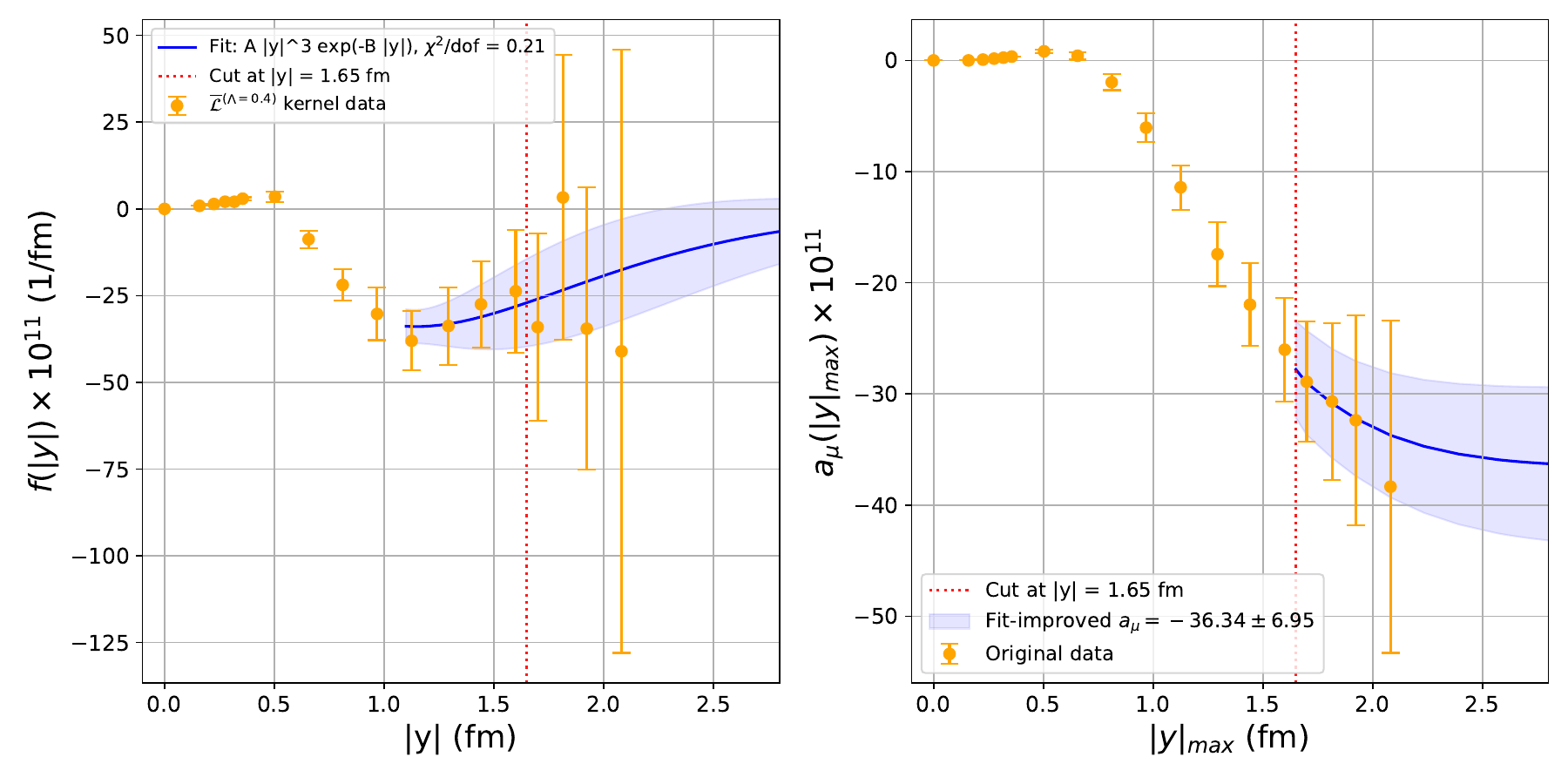}
    \caption{Preliminary results for the light-quark 2+2 disconnected contribution on the ensemble cB64.    The left plot depicts the integrand $f(|y|)$ together with a fit of the form $A~|y|^3~e^{-B|y|}$ for $|y| \gtrsim 1.1$ fm shown as a blue band. The right plot displays the partially integrated $a_\mu(|y|_\text{max})$ with the effect of replacing the data by the fit result for $|y| \geq 1.65$ fm shown as a blue band.}
    \label{fig:light-light-cB}
\end{figure}

In this section we present the preliminary results for the light-quark connected and 2+2 disconnected contributions, evaluated on the coarsest ensemble cB64 with lattice spacing ${a \sim 0.08 \, \text{fm}}$ using 789 configurations. Here we evaluate only the kernel $\overline{\mathcal{L}}^{(\Lambda=0.4)}$ and direction 1111. The results for the light-quark connected contribution are shown in Figure \ref{fig:light-connected-k4}.  With the current statistics, it is possible to resolve the signal for $|y| \lesssim 2$ fm.

In order to control the statistical noise at large distance, we parameterize the data for the integrand with an effective model of the form $A \,|y|^3 e^{-B|y|}$, as suggested in Ref.~\cite{Chao:2021tvp}, where $A$ and $B$ are free parameters to fit the data. This model is motivated from the parameterization of the pole contribution---or in general the single-resonance exchange---at large distances. The model is fitted to the data from a minimum value $|y|_{\text{fit}}$ until the end of the available range; the data are then replaced with the effective model starting from $|y|_{\text{cut}}$. In this preliminary investigation, we choose $|y|_{\text{fit}} \sim 0.79$ fm, corresponding to the peak of the integrand and yielding $\chi^2/\text{d.o.f.} = 0.50$. The right panel of Figure \ref{fig:light-connected-k4} depicts the data together with the effect of their replacement with the effective model after $|y|_{\text{cut}} = 2.0$ fm depicted as a blue band.
In future work, this process will be repeated for a variety of values of $|y|_{\text{fit}}$ and $|y|_{\text{cut}}$ over which an AIC-averaging procedure \cite{Borsanyi:2020mff} can be applied to determine the final extrapolated value for the light-quark connected contribution.

We also present the updated results for the light-quark 2+2 disconnected contribution for the coarsest ensemble cB64.
We follow the same procedure as in the light-quark connected case, fitting and replacing the data in the tail with the same functional form $A\,|y|^3 e^{-B|y|}$, where $A$ and $B$ are independently obtained in this case.
The left panel of Figure \ref{fig:light-light-cB} depicts a representative fit with the fit range beginning at the peak $|y|_{\text{fit}} \sim 1.1$ fm, yielding $\chi^2/\text{d.o.f.} = 0.21$. The data are replaced with the effective model starting from $|y|_{\text{cut}} = 1.65$ fm. The right panel depicts the partially integrated results obtained directly from the data together with the improved estimates obtained using the effective model (shown again as a blue band). 
In Table~\ref{table:light_quarks} we summarize the preliminary results for the light-quark contributions. 
We obtain the total by adding the two light-quark contributions, and the combined total uncertainty is estimated conservatively as a linear sum of the individual errors. 

\begin{table}[t]
    \centering
    \begin{tabular}{lccc}
        \toprule
        Contribution & Diagrams & $\bar{\mathcal{L}}^{(\Lambda=0.4)} \times 10^{11}$ \\
        \midrule
        Light (cB64) & Connected & $128(11)$ \\
        Light (cB64) & Disconnected (2+2) & $-36(7)$ \\
        \midrule
        Light (cB64) & Total & $92(18)$ \\        
        \bottomrule
    \end{tabular}
    \caption{
        Preliminary results for the light-quark connected and 2+2 disconnected contributions and their total, with the total uncertainty obtained conservatively by adding the individual uncertainties linearly, assuming the errors to be 100\% correlated.
    }
    \label{table:light_quarks}
\end{table}

\section{Conclusions and outlook}

We have presented preliminary results for the HLbL contribution to the muon anomalous magnetic moment using the ETMC $N_f = 2+1+1$ twisted-mass gauge ensembles at the physical point. They represent an expansion of ETMC's previous results based on the pseudoscalar pole contributions \cite{ExtendedTwistedMass:2022ofm,ExtendedTwistedMass:2023hin}.

The charm- and strange-quark connected contributions have been extrapolated to the continuum limit using four lattices spacings, including
a systematic error from variations of the fit models and using an AIC procedure. 
The charm-quark connected contribution is still under scrutiny in order to potentially suppress the large lattice artifacts at short distances.

For the light-quark contributions (connected and 2+2 disconnected), we have presented measurements at one lattice spacing only. The currently obtained uncertainty is $\sim 20\%$. By further controlling the long-distance tail with additional statistics, we expect to reduce this uncertainty to $\sim 10\%$.
Simultaneously, a full accounting of systematic uncertainty introduced by the effective model for the long-distance tail is under way.
Calculations of the heavy-flavor and flavor-mixing 2+2 disconnected contributions are under way. Combined with future evaluations of the light-flavor contributions at finer lattice spacings, the present work is a significant step towards a fully controlled, continuum estimate of $a_\mu^{\text{{HLbL}}}$ using the twisted-mass discretization.

\acknowledgments

We thank Antoine G\'{e}rardin and Harvey Meyer for helpful discussions. We are also grateful to the rest of the ETM collaboration for support of this work. This work is supported by the Swiss National Science Foundation (SNSF) through grant No.~200020$\_$208222 and No.~200021-232288. We gratefully acknowledge computing time granted on Piz Daint at Centro Svizzero di Calcolo Scientifico (CSCS) under project {\tt s1197} and {\tt ch15}. We gratefully acknowledge the Gauss Centre for Supercomputing e.V. (www.gauss-centre.eu) for supporting this work through computing time on the GCS supercomputer JUWELS Booster at the Jülich Supercomputing Centre under project {\tt HALISCA}. Evaluations of the QED kernel were possible by using the KQED library~\cite{Asmussen:2022oql}. Ensemble production and measurements for this analysis made use
of tmLQCD~\cite{Jansen:2009xp, Deuzeman:2013xaa, Abdel-Rehim:2013wba, Kostrzewa:2022hsv} and QUDA~\cite{Clark:2009wm, Babich:2011np, Clark:2016rdz}. The figures were produced using matplotlib~\cite{4160265}.

\bibliographystyle{JHEP}
\bibliography{main} 

\end{document}